\documentclass{article}
\usepackage[utf8]{inputenc}
\usepackage{amsmath}
\usepackage{amsthm}
\usepackage{amssymb}
\usepackage{bbm}
\usepackage[round]{natbib}
\usepackage[english]{babel} 
\usepackage{enumerate}
\usepackage[margin=1in]{geometry}
\usepackage{graphicx}
\usepackage{pdfpages}
\usepackage{makecell}
\usepackage{cases}
\usepackage{caption}
\usepackage{subcaption}
\usepackage[normalem]{ulem}
\usepackage{hyperref}
\usepackage{tikz} 

\usepackage{float}
\floatstyle{plaintop}
\restylefloat{table}

\hypersetup{
    colorlinks=false, 
    linktoc=all,     
    linkcolor=blue,  
    hidelinks,
}

\DeclareMathOperator{\rank}{rank}
\DeclareMathOperator{\diag}{diag}

\DeclareMathOperator{\Exp}{E}
\DeclareMathOperator{\logit}{logit}

\newtheorem*{theorem*}{Theorem} 

\definecolor{orangey}{RGB}{210,0,0}

\begin{document}
\title{\vspace{-10mm} \textbf{\Large{Improving the Hosmer-Lemeshow Goodness-of-Fit Test in Large Models with Replicated Trials}}}
\author{\textbf{\normalsize{Nikola Surjanovic and Thomas M. Loughin}}}
\date{\normalsize{Department of Statistics and Actuarial Science, Simon Fraser University, Burnaby, British Columbia V5A 1S6, Canada}}
\maketitle
\baselineskip=14pt 

\section*{\large{Summary}}
{\small The Hosmer-Lemeshow (HL) test is a commonly used global goodness-of-fit (GOF) test that assesses the quality of the overall fit of a logistic regression model. In this paper, we give results from simulations showing that the type 1 error rate (and hence power) of the HL test decreases as model complexity grows, provided that the sample size remains fixed and binary replicates are present in the data. We demonstrate that the generalized version of the HL test by \cite{surjanovic2020generalized} can offer some protection against this power loss. We conclude with a brief discussion explaining the behaviour of the HL test, along with some guidance on how to choose between the two tests.

\noindent \textit{Key Words: } chi-squared test, generalized linear model, goodness-of-fit test, Hosmer-Lemeshow test, logistic regression}

\bigskip \noindent \textcolor{orangey}{\textbf{Note:} The latest version of this 
article is available at \url{https://doi.org/10.1080/02664763.2023.2272223}.}

\section{Introduction}
Logistic regression models have gained a considerable amount of attention as a tool for estimating the probability of success of a binary response variable, conditioning on several explanatory variables. Researchers in health and medicine have used these models in a wide range of applications. One of many examples includes estimating the probability of hospital mortality for patients in intensive care units as a function of various covariates \citep{lemeshow1988predicting}.

Regardless of the application, it is always desirable to construct a model that fits the observed data well. One of several ways of assessing the quality of the fit of a model is with goodness-of-fit (GOF) tests \citep{bilder2014analysis}. In general, GOF tests examine a null hypothesis that the structure of the fitted model is correct. They may additionally identify specific alternative models or deviations to test against, but this is not required. Global (omnibus) GOF tests are useful tools that allow one to assess the validity of a model without restricting the alternative hypothesis to a specific type of deviation. 

An example of a well-known global GOF test for logistic regression is the Hosmer-Lemeshow (HL) test, introduced by \cite{hosmer1980goodness}. The test statistic is a Pearson statistic that compares observed and expected event counts from data grouped according to ordered fitted values from the model. The decision rule for the test is based on comparing the test statistic to a chi-squared distribution with degrees of freedom that depend on the number of groups used to create the test statistic. The HL test is relatively easy to implement in statistical software, and since its creation, the HL test has become quite popular, particularly in fields such as biostatistics and the health sciences. 

Despite its popularity, the HL test is known to have some drawbacks. In both experimental and observational studies, it is possible to have data for which binary observations have the same explanatory variable patterns (EVPs). In this case, responses can be aggregated into binomial counts and trials. When there are observations with the same EVP present in the data, it is possible to obtain many different p-values depending on how the data are sorted \citep{bertolini2000one}. A related disadvantage of HL-type tests is that the test statistic depends on the way in which groups are formed, as remarked upon by \cite{hosmer1997comparison}. In this paper we will highlight a related but different problem with the HL test that does not appear to be well known.

Models for logistic regression with binary responses normally assume a Bernoulli model where the probability parameter is related to explanatory variables through a logit link. As mentioned, when several observations have the same EVP, responses can be summed into binomial counts. Other times, the joint distribution of covariates may cause observed values to be clustered into near-replicates, so that the responses might be viewed as being approximately binomially distributed. These cases present no problem for model fitting and estimating probabilities. However, it turns out that this clustering in the covariate space may materially impact the validity of the HL test applied to the fitted models.

For such data structures, a chi-squared distribution does not represent the null distribution of the HL test statistic well in finite samples, as suggested by simulation results in Section \ref{sec:sim_results}. This, in turn, adversely affects both the type 1 error rate and the power of the HL test. \cite{bertolini2000one} demonstrated that it is possible to obtain a wide variety of p-values and test statistics when there are replicates in the data, simply by reordering the data. Our analysis also deals with replicates in the data. However, we find a different phenomenon: as the model size grows for a fixed sample size, the type 1 error rate tends to decrease.

In this paper we show that the HL test can be improved upon by using another existing global GOF test, the generalized HL (GHL) test from \cite{surjanovic2020generalized}. Empirical results suggest that the GHL test performs reasonably well even when applied to moderately large models fit on data with exact replicates or clusters in the covariate space. We offer a brief discussion as to why one might expect clustering in the covariate space to affect the regular HL test. A simple decision tree is offered to summarize when each test is most appropriate.

An overview of the HL and GHL tests is given in Section \ref{sec:methods}. The design of the simulation study comparing the performance of these two tests is outlined in Section \ref{sec:sim_design}, with the results given in Section \ref{sec:sim_results}. We end with a discussion of the implications of these results and offer some guidance on how to choose between the two tests in Section \ref{sec:discussion}.


\section{Methods}
\label{sec:methods}
In what follows, we use the notation of \cite{surjanovic2020generalized}. We let $Y \in \{0,1\}$ denote a binary response variable associated with a $d$-dimensional covariate vector, $X \in \mathbb{R}^d$, where the first element of $X$ is equal to one. The pairs $(X_i, Y_i)$, $i=1,\ldots,n$ denote a random sample, with each pair being distributed according to the joint distribution of $(X,Y)$. The observed values of $(X_i, Y_i)$ are denoted using lowercase letters as $(x_i, y_i)$. 

In a logistic regression model with binary responses, one assumes that
$$\Exp(Y|X=x) = \pi(\beta^\top x) = \frac{\exp(\beta^\top x)}{1+\exp(\beta^\top x)},$$
for some $\beta \in \mathbb{R}^d$. The likelihood function is
$$\mathcal{L}(\beta) = \prod_{i=1}^n \pi(\beta^\top x_i)^{y_i} (1-\pi(\beta^\top x_i))^{1-y_i}.$$
From this likelihood, a maximum likelihood estimate (MLE), $\beta_n$, of $\beta$ is obtained. 

\bigskip \noindent \textbf{The HL Test Statistic} \\
To compute the HL test statistic, one partitions the observed data, $(x_i, y_i)$, into $G$ groups. Typically, the groups are created so that fitted values are similar within each group and the groups are approximately of equal size. To achieve this, a partition is defined by a collection of $G+1$ interval endpoints, $-\infty = k_0 < k_1 < \cdots < k_{G-1} < k_G = \infty$. The $k_g$ often depend on the data, usually being set equal to the logits of equally-spaced quantiles of the fitted values, $\hat\pi_i = \pi(\beta_n^\top x_i)$. We define $I_i^{(g)} = \mathbbm{1}(k_{g-1} < \beta_n^\top x_i \leq k_g)$, $O_g = \sum_{i=1}^n y_i I_i^{(g)}$, $E_g = \sum_{i=1}^n \hat\pi_i I_i^{(g)}$, $n_g = \sum_{i=1}^n I_i^{(g)}$, and $\bar{\pi}_g = E_g/n_g$, where $\mathbbm{1}(A)$ is the indicator function on a set $A$. With this notation, the number of observations in the $g$th group is represented by $n_g$, and $\bar{\pi}_g$ denotes the mean of the fitted values in this group. The HL test statistic is a quadratic form that is commonly written in summation form as 
$$\widehat{C}_G = \sum_{g=1}^G \frac{(O_g - E_g)^2}{n_g \bar{\pi}_g (1-\bar{\pi}_g)}.$$

When $G>d$, \cite{hosmer1980goodness} find that the HL test statistic is asymptotically distributed as a weighted sum of chi-squared random variables under the null hypothesis, after checking certain conditions of Theorem 5.1 in \cite{moore1975unified}. Precisely,
\begin{equation}
\label{eq:HL_limit}
    \widehat{C}_G \xrightarrow{d} \chi^2_{G-d} + \sum_{j=1}^d \lambda_j \chi_{1j}^2,
\end{equation}
with each $\chi_{1j}^2$ being a chi-squared random variable with 1 degree of freedom, and each $\lambda_j$ an eigenvalue of a certain matrix that depends on both the distribution of $X$ and the vector $\beta_0$, the true value of $\beta$ under the null hypothesis. \cite{hosmer1980goodness} conclude through simulations that the right side of (\ref{eq:HL_limit}) is well approximated by a $\chi^2_{G-2}$ distribution in various settings.

The HL test statistic and the corresponding p-value both depend on the chosen number of groups, $G$. Typically, $G=10$ groups are used, so that observations are partitioned into groups that are associated with ``deciles of risk''. Throughout this paper we use 10 groups, and therefore compare the HL test statistic to a chi-squared distribution with 8 degrees of freedom, but the results hold for more general choices of $G$.

\bigskip \noindent \textbf{The GHL Test Statistic} \\
The GHL test introduced by \cite{surjanovic2020generalized} generalizes several GOF tests, allowing them to be applied to other generalized linear models. Tests that are generalized by the GHL test include the HL test \citep{hosmer1980goodness}, the Tsiatis \citep{tsiatis1980note} and generalized Tsiatis tests \citep{canary2016summary}, and a version of the ``full grouped chi-square'' from \cite{hosmer2002goodness} with all weights equal to one. The test statistic is a quadratic form like $\widehat{C}_G$, but with important changes to the central matrix. The theory behind this test depends on the residual process, $R_n^1(u)$, $u \in \mathbb{R}$, defined in \cite{stute2002model}. In the case of logistic regression,
$$ R_n^1(u) = \frac{1}{\sqrt{n}} \sum_{i=1}^n [Y_i - \pi(\beta_n^\top X_i)] \mathbbm{1}(\beta_n^\top X_i \leq u) , $$
a cumulative sum of residuals that are ordered according to the size of their corresponding fitted values. This process is transformed into a $G$-dimensional vector, $S_n^1$, which forms the basis of the HL and GHL test statistics, with 
$$ S_n^1 = (R_n^1(k_1) - R_n^1(k_0), \ldots, R_n^1(k_G) - R_n^1(k_{G-1}) )^\top. $$

In order to approximate the variance of $S_n^1$, we need to define several matrices. Let
\begin{align*}
    \left(G_n^{*}\right)_{gi} &= I_i^{(g)}, \\
    V^{*1/2} &= \diag\left( [\pi(\beta_0^\top x_i) (1-\pi(\beta_0^\top x_i))]^{1/2} \right),
\end{align*}
for $i = 1, \ldots, n$, and $g = 1, \ldots, G$. Also, define $X^*$ to be the $n \times d$ matrix with $i$th row given by $x_i^\top$, and let $V_n^{*1/2}$ be the same as $V^{*1/2}$, but evaluated at the estimate $\beta_n$ of $\beta_0$. Finally, define 
\begin{align}
\label{eq:Sigma_n}
    \Sigma_n 
      &= \frac{1}{n} G_n^* \left(V_n^* - V_n^* X^* (X^{*\top} V_n^* X^*)^{-1} X^{*\top} V_n^{*}\right) G_n^{*\top} \nonumber \\
      &= \frac{1}{n} G_n^* V_n^{*1/2} \left(I_n - V_n^{*1/2} X^* (X^{*\top} V_n^* X^*)^{-1} X^{*\top} V_n^{*1/2}\right) V_n^{*1/2} G_n^{*\top},
\end{align}
where $I_n$ is the $n \times n$ identity matrix. 

For logistic regression models, the GHL test statistic is then
$$ X^2_{\text{GHL}} = S_n^{1\top} \Sigma_n^{+} S_n^1, $$
where $\Sigma_n^{+}$ is the Moore-Penrose pseudoinverse of $\Sigma_n$. Under certain conditions given by \cite{surjanovic2020generalized}, 
$$ S_n^{1\top} \Sigma_n^{+} S_n^1 \xrightarrow{d} \chi^2_\nu, $$
where $\nu = \rank(\Sigma)$, with $\Sigma$ a matrix defined in their paper. Since the rank of $\Sigma$ might be unknown, they use the rank of $\Sigma_n$ as an estimate. We use the same approach to estimating $\nu$, empirically finding that the estimated rank of $\Sigma_n$ is often equal to $G-1$ for logistic regression models.

The GHL test statistic for logistic regression models is equivalent to the Tsiatis GOF test statistic \citep{tsiatis1980note} and the $X^2_w$ statistic from \cite{hosmer2002goodness} with all weights set equal to 1, when a particular grouping method is used---that is, when $G_n^*$ is the same for all methods. However, use of the GHL test is justified for a wide variety of GLMs and grouping procedures with a proper modification of $\Sigma_n$, as described by \cite{surjanovic2020generalized}. For both the HL and GHL tests, we use the grouping method proposed by \cite{surjanovic2020generalized}, which uses random interval endpoints, $k_g$, so that $\sum_{i=1}^n \hat\pi_i (1-\hat\pi_i) I_i^{(g)}$ is roughly equal between groups. Further details of the implementation are provided in the supplementary material of their paper.

It is important to note that $\Sigma_n$ is a non-diagonal matrix that standardizes and accounts for correlations between the grouped residuals in the vector $S_n^1$. This can be seen from (\ref{eq:Sigma_n}), which shows that $\Sigma_n$ contains a generalized hat matrix for logistic regression. In contrast, when written as a quadratic form, the central matrix of the HL test statistic is diagonal and does not account for the number of parameters in the model, $d$, when standardizing the grouped residuals. We expect this standardization to be very important when exact replicates are present, as the binomial responses might be more influential than sparse, individual binary responses.

It is extremely common to fit logistic regression models to data where multiple Bernoulli trials are observed at some or all EVPs, even when the underlying explanatory variables are continuous. As with any fitted model, a test of model fit would be appropriate, and the HL test would likely be a candidate in a typical problem. It is therefore important to explore how the HL and GHL tests behave with large models when exact or near-replicates are present in the data.


\section{Simulation Study Design}
\label{sec:sim_design}
We compare the performance of the HL and GHL tests by performing a simulation study. Of particular interest is the rejection rate under the null, when the tests are applied to moderately large models that are fit to data with clusters or exact replicates in the covariate space.

In all settings, the true regression model is 
\begin{equation}
\label{eq:sim_prob}
    \Exp(Y|X=x) = \logit(\beta_0 + \beta_1 x_1 + \ldots + \beta_{d-1} x_{d-1}),
\end{equation}
with $d$ in $\{2, 3, \ldots, 25\}$. Here, $\beta_0$ represents the intercept term. To produce replicates in the covariate space, $m \leq n$ unique EVPs are drawn randomly from a $(d-1)$-dimensional spherical normal distribution with marginal mean 0 and marginal variance $\sigma^2=1$ for each simulation realization. At each EVP, $n/m$ replicate Bernoulli trials are then created, with probabilities determined by (\ref{eq:sim_prob}). In our simulation study, we fix $n=500$ and select $m \in \{50, 100, 500\}$ so that the number of replicates at each EVP, $n/m$, is 10, 5, or 1, respectively. 

We set $\beta_0 = 0.1$ and $\beta_1 = \ldots = \beta_{d-1} = 0.535 / \sqrt{d-1}$. This results in fitted values that rarely fall outside the interval $[0.1, 0.9]$, regardless of the number of parameters in the model, so that the number of expected counts in each group is sufficiently large for the use of the Pearson-based test statistics. 

We also perform some simulations with $n=100$, using smaller values of $d$ and $m$ than for $n=500$. However, we focus on results for $n=500$ because we are then able to increase the number of replicates per EVP, $n/m$, while still maintaining large enough $m$ so that it is possible to create ten viable groupings. In each simulation setting, 10,000 realizations are produced. All simulations are performed using \textsf{R}.


\section{Simulation Results}
\label{sec:sim_results}
\begin{figure}[!htb]
\centering
\begin{subfigure}{.49\textwidth}
  \centering
  \includegraphics[width=\linewidth]{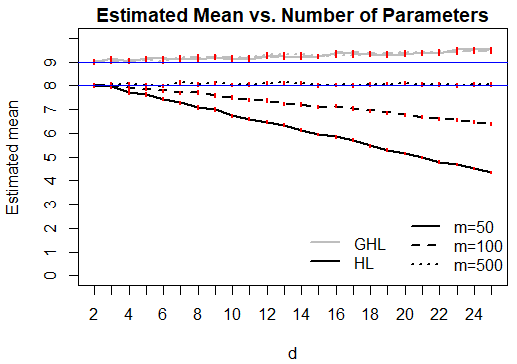}
  \label{fig:mean}
\end{subfigure}
\begin{subfigure}{.49\textwidth}
  \centering
  \includegraphics[width=\linewidth]{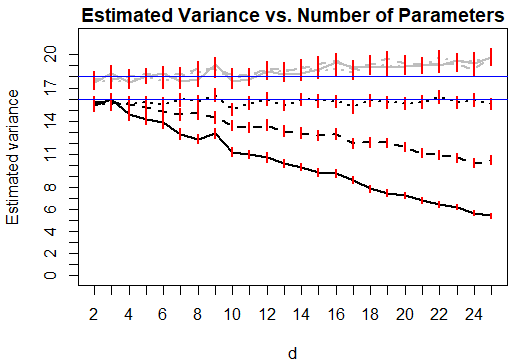}
  \label{fig:variance}
\end{subfigure}
\begin{subfigure}{.49\textwidth}
  \centering
  \includegraphics[width=\linewidth]{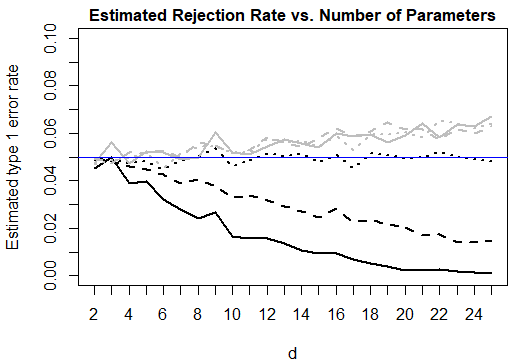}
  \label{fig:type1}
\end{subfigure}
\caption{Null simulation results. Solid red lines are approximate 95\% CIs. Intervals are omitted for the type 1 error rate plot, but can be approximated by adding and subtracting 0.005 from the estimated rejection rate.}
\label{fig:null_results}
\end{figure}

Figure \ref{fig:null_results} presents plots of the sample mean and variance of the HL and GHL test statistics against the number of variables in the model, separately for each $m$. An analogous plot of the estimated type 1 error rate of the tests against the number of variables is also presented. For the HL test, all three statistics show a clear decreasing pattern with increasing model size when replicates are present, with a sharper decrease when the number of replicates per EVP is larger. Since the estimated variance is not always twice the size of the mean, we can infer that the chi-squared approximation to the null distribution of the HL test statistic is not adequate in finite samples for these data structures. Simulation results with a sample size of $n=100$ are not displayed, but are quite similar.

From the same figure, we see that the GHL test performs well in the settings considered. The estimated mean and variance of the test statistic stay close to the desired values of $G-1=9$ and $2(G-1)=18$. We note that the GHL test can have an inflated type 1 error rate, particularly when it is applied to highly complex models. The models considered here are only moderately large, with $d \leq \min \{ n/20, m/2 \}$. If one wishes to use the GHL test to assess the quality of fit of larger models with only a moderate sample size, one should be wary of an inflated type 1 error rate that can become considerably large for complex models. A possible explanation for this is that estimating the off-diagonal elements of the matrix $\Sigma_n$ can potentially introduce a considerable amount of variance into the test statistic in small samples.

Recall from (\ref{eq:HL_limit}) that the asymptotic $\chi^2_{G-2}$ distribution for the HL test proposed by \cite{hosmer1980goodness} is based on a sum of chi-squares, where one has $G-d$ degrees of freedom. We investigated whether maintaining $G=10$ while increasing $d$ contributes to the phenomena we have observed. We set $G=26$ and performed a similar simulation study. The adverse behavior of the HL statistic still persists despite this modification. 

\begin{figure}[!htb]
\centering
\begin{subfigure}{.49\textwidth}
  \centering
  \includegraphics[width=\linewidth]{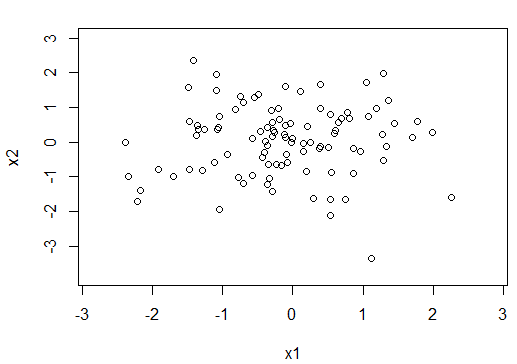}
  \label{fig:sub1}
\end{subfigure}
\begin{subfigure}{.49\textwidth}
  \centering
  \includegraphics[width=\linewidth]{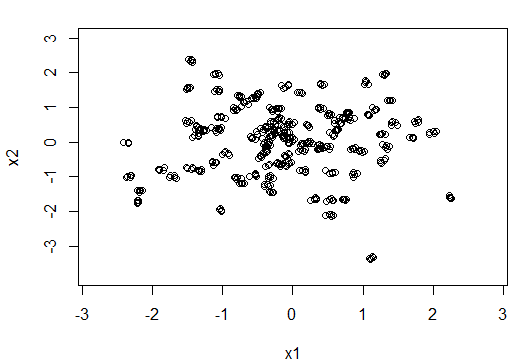}
  \label{fig:sub2}
\end{subfigure}
\begin{subfigure}{.49\textwidth}
  \centering
  \includegraphics[width=\linewidth]{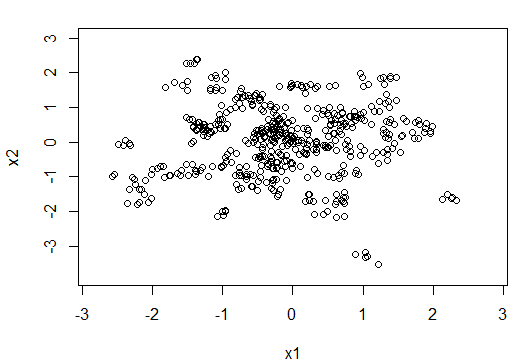}
  \label{fig:sub3}
\end{subfigure}
\begin{subfigure}{.49\textwidth}
  \centering
  \includegraphics[width=\linewidth]{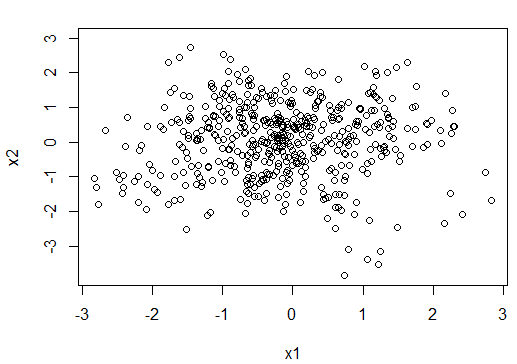}
  \label{fig:sub4}
\end{subfigure}
\caption{Example of clustering in the covariate space with two predictor variables, $X_1$ and $X_2$. Top left to bottom right: $\sigma^2_e = 0$, $0.001$, $0.01$, and $0.1$, with $\sigma^2=1$.}
\label{fig:clustering}
\end{figure}

We also investigated the effect of near-replicate clustering in the covariate space. We fixed $n$ and $m$ as in Section \ref{sec:sim_design}, but added a small amount of random noise with marginal variance $\sigma^2_e$ to each replicate within the $m$ sampled vectors. The amount of clustering was controlled by varying $\sigma^2_e$, as shown in Figure \ref{fig:clustering}. As expected, increasing $\sigma^2_e$ reduces the severity of the decreasing mean, variance and type 1 error rate for the HL test statistic. However, the pattern remains evident while $\sigma^2_e/\sigma^2$ remains small.


\section{Discussion}
\label{sec:discussion}
The original HL test, developed by \cite{hosmer1980goodness}, is a commonly-used test for logistic regression among researchers in biostatistics and the health sciences. Although its performance is well documented \citep{lemeshow1982review, hosmer1997comparison, hosmer2002goodness}, we have identified an issue that does not seem to be well known. For moderately large logistic regression models fit to data with clusters or exact replicates in the covariate space, the null distribution of the HL test statistic can fail to be adequately represented by a chi-squared distribution in finite samples. Using the original chi-squared distribution with $G-2$ degrees of freedom can result in a reduced type 1 error rate, and hence lower power to detect model misspecifications. Based on the results of the simulation study, the GHL test can perform noticeably better in such settings, albeit with a potentially inflated type 1 error rate.

Similar behaviour of the HL test was observed in \cite{surjanovic2020generalized}, where the regular HL test was naively generalized to allow for it to be used with Poisson regression models. In their setup, even without the presence of clusters or exact replicates in the covariate space, as the number of model parameters increased for a fixed sample size, the estimated type 1 error rate decreased. The central matrix in the GHL test statistic, $\Sigma_n$, makes a form of correction to the HL test statistic by standardizing and by accounting for correlations between the grouped residuals that comprise the quadratic form in both the HL and GHL tests. This is evident from (\ref{eq:Sigma_n}), which shows that $\Sigma_n$ contains the generalized hat matrix subtracted from an identity matrix. To empirically assess the behaviour of $\Sigma_n$, we varied $\sigma^2_e$ in the setup with replicates and added noise, described at the end of Section~\ref{sec:sim_results}. For large $d$ and moderate $n$, both fixed, we found that the diagonal elements of $\Sigma_n$ tend to shrink, on average, as $\sigma^2_e$ decreases. In contrast, the elements of the HL central matrix remain roughly constant. Therefore, the GHL statistic seems to adapt to clustering or replicates in $X$, whereas the HL test statistic does not.

In logistic regression with exact replicates, grouped binary responses can be viewed as binomial responses that can be more influential. In this scenario, as $d$ increases for a fixed sample size $n$, the distribution of the regular HL test statistic diverges from a single chi-squared distribution, suggesting that the standardization offered by the central GHL matrix becomes increasingly important. 

The real-life implications of the reduced type 1 error rate and power of the regular HL test are that in models with a considerable number of variables---provided that the data contains clusters or exact replicates---the HL test has limited ability to detect model misspecifications. Failure to detect model misspecification can result in retaining an inadequate model, which is arguably worse than rejecting an adequate model due to an inflated type 1 error rate, particularly when logistic regression models are used to estimate probabilities from which life-and-death decisions might be made.

\begin{figure}
\centering
\begin{tikzpicture}[
  sibling distance=20em,
  every node/.style = {
    draw, 
    align=center,
    shape=rectangle
  }]
\node {\begin{tabular}{c} Is the model small or moderate in size? \\ ($d \leq \min \{n/20, m/2 \}$) \end{tabular}}
    child { node {\begin{tabular}{c} Are there replicates or \\ clusters of observations? \end{tabular}}
      child { node {Use HL} }
      child { node[xshift=-8em] {\begin{tabular}{c} Try both tests, \\ but proceed with caution \end{tabular}} }
    } 
    child { node {Very large $n$?} 
      child { node[xshift=8em] {\begin{tabular}{c} Are there replicates or \\ clusters of observations? \end{tabular}} 
        child { node[xshift=8em] {Use HL} }
        child { node {Use GHL or both tests} }
      }
      child { node {Use GHL} }
    };
\end{tikzpicture}
\caption{Decision tree offering guidance on how to choose between the two GOF tests when $G=10$ and $d \lesssim 25$. In each decision, left=no and right=yes.}
\label{fig:decision_tree}
\end{figure}
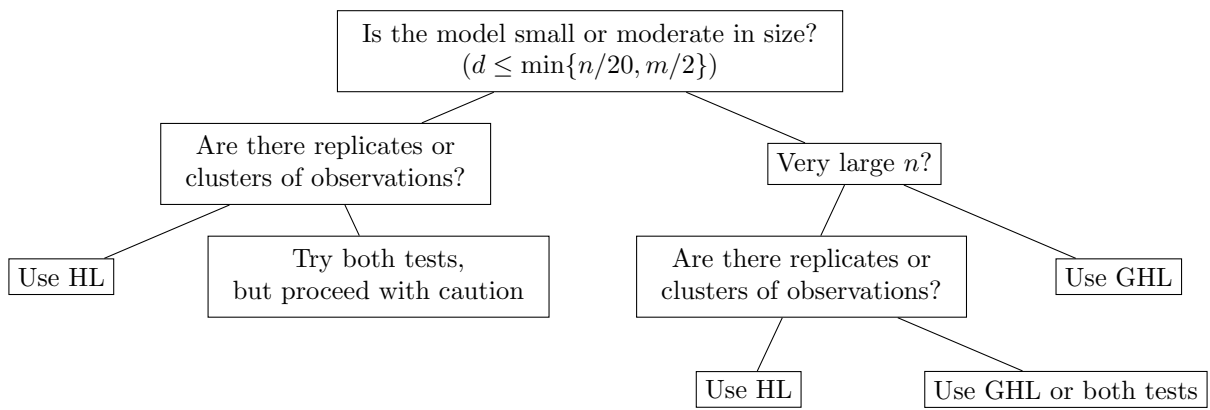

Our advice for choosing between the two GOF tests is displayed as a simple decision tree in Figure~\ref{fig:decision_tree}. The advice should be interpreted for $G=10$ groups, the most commonly used number of groups. With large samples, provided that $m$ is sufficiently large compared to $d$, it should generally be safe to use the GHL test. Our simulations explored models with $d \leq 25$, so some caution should be exercised if the GHL test is to be used with larger models. For small or moderate samples, such as when $n=100$ or $500$, it is important to identify whether there are clusters or exact replicates in the covariate space. One can compute the number of unique EVPs, $m$, and compare this number to the sample size, $n$. If $n/m \geq 5$, say, then there is a considerable amount of ``clustering''. For data without exact replicates, clusters can still be detected using one of many existing clustering algorithms, and the average distances between and within clusters can be compared. Informal plots of the $x_i$ projected onto a two- or three-dimensional space can also be used as an aid in this process. 

If there is no evidence of clustering or replicates, the HL test should not be disturbed by this phenomenon. On the other hand, if there is a noticeable amount of clustering, and the regression model is not too large, say $d \leq \min \{ n/20, m/2 \}$, where $m$ also represents the number of estimated clusters, then one can use the GHL test. In the worst-case scenario with a small sample size, clustering, and a large regression model, one can use both tests as an informal aid in assessing the quality of the fit of the model, recognizing that GHL may overstate the lack of fit, while HL may understate it. If the two tests agree, then this suggests that the decision is not influenced by the properties of the tests. When they disagree, conclusions should be drawn more tentatively.

\section*{Acknowledgements}
We acknowledge the support of the Natural Sciences and Engineering Research Council of Canada (NSERC), [funding reference number RGPIN-2018-04868]. Cette recherche a été financée par le Conseil de recherches en sciences naturelles et en génie du Canada (CRSNG), [numéro de référence RGPIN-2018-04868].

\setlength{\bibsep}{0pt plus 0.3ex} 
\bibliography{main.bib}
\bibliographystyle{plainnat}

\clearpage
\newpage

\end{document}